# A Statistical Model for Ideal Team Selection for A National Cricket Squad

Sadia Tasnim Swarna [1*], Shamim Ehsan [1] and MD. Saiful Islam [1]

[1] Department of Computer Science and Engineering, Shahjalal University of Science & Technology, Sylhet, Bangladesh.

Email: sadiatasnimswarna@gmail.com, ehsanhrid@gmail.com, saiful-cse@sust.edu

**Keywords:**

- *Statistical analysis, cricket, sports, neural network*
- *SUST, ICERIE*

**Abstract**: Cricket is a game played between two teams which consists of eleven players each. Nowadays cricket game is becoming more and more popular in Bangladesh and other South Asian Countries. Before a match people are very enthusiastic about team squads and "Which players are playing today?", "How well will MR. X perform today?" are the million dollar questions before a big match. This article will propose a method using statistical data analysis for recommending a national team squad. Recent match scorecards for domestic and international matches played by a specific team in recent years are used to recommend the ideal squad. Impact point or rating points of all players in different conditions are calculated and the best ones from different categories are chosen to form optimal line-ups. To evaluate the efficiency of impact point system, it will be tested with real time match data to see how much accuracy it gives.

## 1. INTRODUCTION

Cricket is a sport played by two teams containing 11 players in each side. It has become so popular that it is an integral part of the South Asian culture. The exponential growth of popularity that cricket can bring to any individuals, and in the days of IPL and other T20 multi-million cricket leagues, the game of cricket can give huge financial benefits to the players, so it attracts many aspirants to fight for a chance in the national team. That is why the selection committee has to be very careful about choosing the right squad for a match. They have the tough job of selecting the 15 man squad among hundreds of inform players. We are trying to build a system that should recommend the ideal cricket squad for a national Cricket Team Based on their domestic Cricket league. Though the game of cricket is played in mainly three variants The Test, One Day cricket and Twenty-twenty cricket which is also known as T20 cricket, The format of the game that is discussed in this paper is one day cricket. An ideal squad for Bangladesh cricket team in one day international match is recommended in this paper.

## 2. RELATED WORKS

There have been some researches on predicting outcomes of the game of cricket. A Team from BRAC university (members: Fahad Munir, Md. Kamrul Hasan, Sakib Ahmed Sultan Md. Quraish) (2015) works on T20 game winning prediction while the match is in progress. They used decision tree method for their approach .A paper by Satyam Mukherjee (2012) predicted individual performance in Cricket using Social Network Analysis (SNA. A team from Sam Houston State University (members: Ananda B. W. Manage, Stephen M. Scariano, Cecil R. Hallum) (2013) also predict performance of T20-World Cup Cricket 2012 using ranking method. S.N. OMKAR and R. VERMA (2011) have formulated a system using genetic algorithm. Harsha Perera, Jack Davis and Tim B. Swartz have proposed a system for optimal lineups in twenty-20 cricket using relative value statistics and simulated annealing where a lineup consists of three components: team selection, batting order and bowling order (2011). Pabitra Kr. Dey, Abhijit Banerjee, Dipendra Nath Ghosh, Abhoy Chand Mondal has designed a methodical model based on Analytical hierarchy Process (AHP) (2014) for estimation of player price in IPL. This model gives a systematic way to select the important attributes and calculate the weights based on expert opinion to measure the optimal price for a player which will help the IPL team owner to select the player according their budget and strategies.

---

* Corresponding author: ehsanhrid@gmail.com

## 3. DATA COLLECTION

All data was scraped from www.espncrinfo.com and icc-cricket.com. This site contains all test match and first class match scorecard since 1877, all ODI matches since 1971 onwards. For our paper, all of the matches of Dhaka premier league 2016, 2015 and 2014 are collected from espncricinfo and formatted. As this paper is only for One-day International cricket squad recommendation for Bangladesh team, T-20 domestic leagues like BPL and first class matches are excluded in the dataset.

## 4. METHODOLOGY

An ideal cricket team squad has 15 members. Each player of a team is assigned a role in the game of cricket. We can generalize the roles into four categories:

    1) Specialist batsman
    2) Specialist bowler
    3) All-rounder
    4) Wicket-keeper

Ideal distribution of the players are given below:

Table 1: Ideal Distribution of players in a squad

| Role Type | Quantity |
|---|---|
| Specialist Batsman | 6 |
| Specialist Bowler | 6 |
| Wicket Keeper | 1 |
| All-Rounder | 2 |

Every team relies mostly on their specialist ones, so 6 from each specialist departments are chosen, leaving 2 slot for the all-rounders and one for wicket-keeper. If we look at the squads of ICC cricket world cup 2015, we can see that most of the teams choose 6 batsmen, 6 bowler, 2 genuine allrounders and a wicket-keeper. (Table 2)

Table 2: Role Distribution of all teams in ICC cricket world cup 2015

| Country Name | Specialist Batsman | Specialist Bowler | All rounder | Keeper |
|---|---|---|---|---|
| Afganistan | 6 | 5 | 2 | 2 |
| Australlia | 6 | 6 | 2 | 1 |
| Bangladesh | 6* | 6 | 2 | 1 |
| India | 6* | 6 | 2 | 1 |
| Ireland | 4 | 7 | 2 | 2 |
| England | 5* | 5 | 4 | 1 |
| Pakistan | 6* | 6 | 2 | 1 |
| Srilanka | 6 | 6 | 2 | 1 |
| Zimbabwe | 7 | 4 | 2 | 2 |
| West Indies | 6 | 6 | 2 | 1 |
| South Africa | 6 | 6 | 1 | 2 |
| New Zealand | 5 | 6 | 2 | 2 |

It is clearly seen that most teams prefer the combination of 6 specialist batsmen, 6 specialist bowler, 2 allrounders and a Wicket keeper. It is clearly seen that most teams prefer the combination of 6 specialist batsman, 6 specialist bowler, 2 all-rounders and a Wicket keeper. Asterisk in specialist batsman column indicates there is a part time wicket keeper among them.

For selecting the cricket team squad, various factors should be considered. Different set of input should be provided for each section. If the domain is specialized batsman category, the system should take player's recent performance, past history, average, strike rate, history with opponent team as input. If the player is a specialized bowler, total wickets taken, recent strike rate and economy are considered as input.

*4.1 Bowler Analysis:*

In general, specialist bowlers are of two types- Fast Bowlers and Spin Bowlers. For selecting spin bowlers, we plotted strike rate and economy of the different types of spinners like off break spinner, leg break spinners, slow left arm spinners etc, then rate every bowler a number between 1 and 100 and call the number ranking point of that specific bowler.

Table 3: Top Rating point holder Spinners

| Player Name | Rating Point |
|---|---|
| Shakib Al Hasan | 82.2066 |
| Alok Kapali | 79.6795 |
| Mehedi Hasan | 75.1425 |
| Jubair Hossain | 72.1134 |

For fast or medium fast bowlers, we put 50 on economy, 50 on strike rate and give additional 10 percent point if the player plays any international matches in recent 1 year.

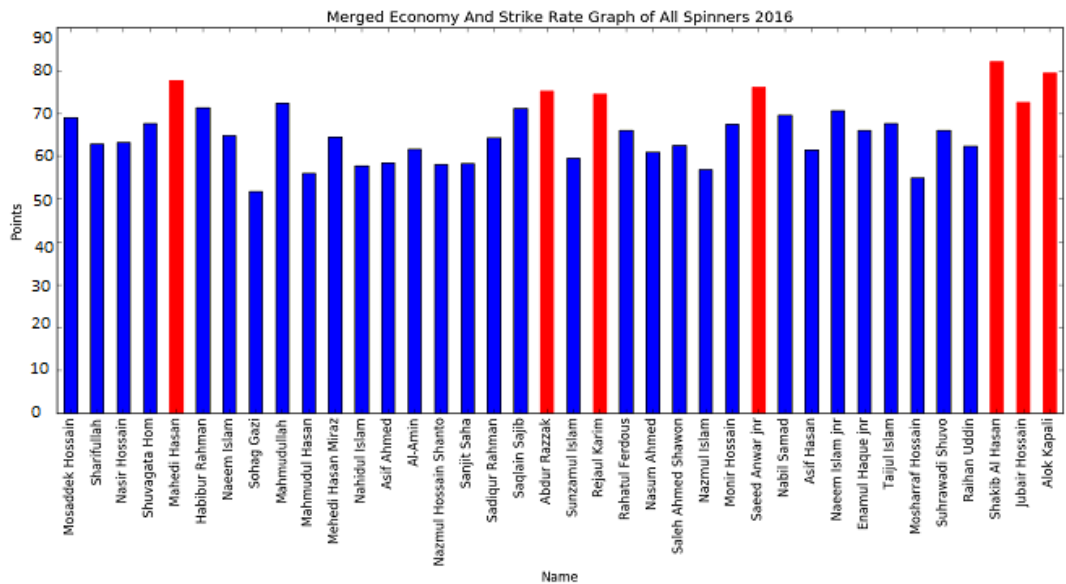

Fig 1: Economy and Strike rate graph for all spinners

Table 4: Top Ranking point holder Pacers

| Player Name | Rating Point |
|---|---|
| Mashrafee Mortaza | 81.92 |
| Taskin Ahmed | 79.78 |
| Al-Amin Hossain | 78.50 |
| Ziaur Rahman | 78.0 |
| S Rana | 76.8 |

*4.1.2 Batsman Analysis:*

Openers, middle order batsman, lower order batsman/pinch hitters - we select the batsmans from these three categories.
*Openers:*
For openers, the ones who open at least one match in Dhaka Premier League 2014-16 or a lower order batsman having an excellent strike rate over 100 are counted.

Table 5: Top rating Point Holder Openers:

| Player Name | Rating Point |
|---|---|
| Shamsur Rahman | 98.66 |
| Tamim Iqbal | 92.28 |
| Imrul Kayes | 92.95 |
| Anamul Haque | 91.1 |

*Middle-Order Batsman:*
For middle order batsman, we put more priority in average rather than strike rate because at the middle stage, it is not that important to score at the strike rate of 100.

Table 6: Top Rating Point of Middle Order Batsman:

| Player Name | Rating Point |
|---|---|
| Salman Hossain | 73.0 |
| Mushfiqur Rahim | 74.0 |
| Tushar Imran | 68.0 |
| Mahmudullah | 68.7 |
| Mosaddek Hossain | 71.0 |
| Nasir Hossain | 61.0 |

*Lower-order-Batsman:*
The 6th and 7th batsman of a batting order generally enters into the crease during the slog overs, where hard hitting is required. So it is more logical to select a batsman who can hit the ball hard, has a higher strike rate rather than selecting someone with a decent average. We rank the lower order batsmans' mostly on their strike rates and number of boundaries they have scored. If the batsman is a part time bowler, additional points are added to his account. Top Rating point holder lower-order Batsman list is shown in Table 7

Table 7: Top Rating Point for Lower Order batsman:

| Player Name | Rating Point |
|---|---|
| Nasir Hossain | 74.05 |
| Sabbir Rahman | 82.88 |
| Shuvagata Home | 90.833 |

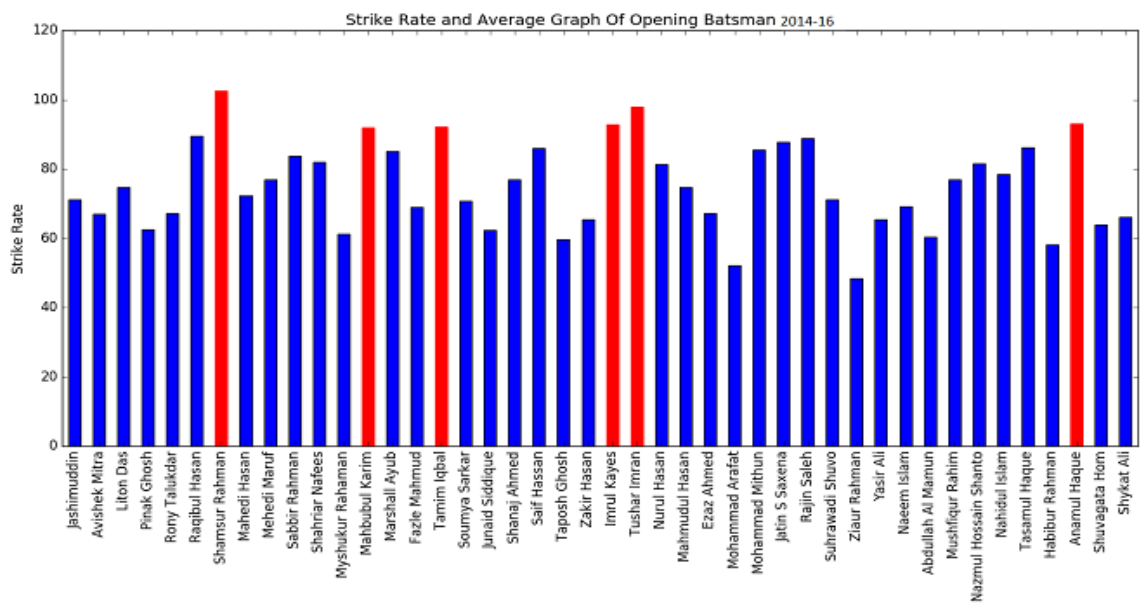

Fig 2: Strike Rate and average of openers

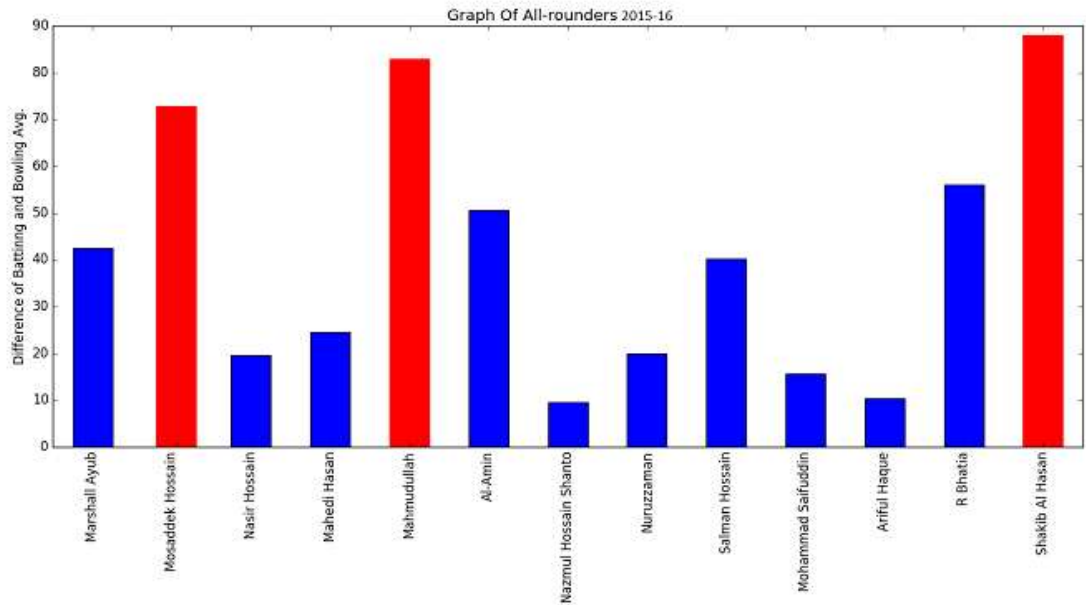

Fig 3: Difference of batting and bowling average of all rounders

*4.1.3 All-rounder analysis:*
The most formal method to compare between two all-rounder is used here. Difference between batting and bowling average of an individual player is used as key factor of ranking all-rounders. The higher the difference is, the better the player. Then the difference is converted to 100 as our other rating points are calculated in 100 also. Our number one all-rounder Shakib Al-Hasan tops our list, but Mahmudullah and young-star Mosaddek Hossain gives him a tough fight. Top Rating Point holder All-rounders list is shown in Table 8

Table 8: Top Rating Point holder All-Rounders:

| Player Name | Rating Point |
|---|---|
| Shakib Al-Hasan | 88.0 |
| Mahmudullah | 82.88 |
| Al-Amin | 50.58 |
| Mosaddek Hossain | 76 |

## 5.CONCLUSION

*5.1 RESULT AND ANALYSIS:*
So our recommended squad based on the statistical analysis on domestic league cricket 2013-16 and the squad of last ODI series Bangladesh have played both is shown on Table 9 along with the squad of last ODI series Bangladesh have played. From the table we can see that there are 6 miss-matches. Two of our best young sensations Mustifizur Rahman and Soumya Sarkar haven't played much in the dhaka premirer league, but considering their recent performance, they are included in our recommended squad.

Table 9: Recommended and current squad

| Recommended Squad | Rating Point | Current Squad |
|---|---|---|
| Shamsur Rahman | 98.66 | Soumya Sarkar |
| Tamim Iqbal | 92.28 | Tamim Iqbal |
| Imrul Kayes | 92.95 | Imrul Kayes |
| Salman Hossain | 73.0 | Liton Kumar Das |
| Mushfiqur Rahim | 74.0 | Mushfiqur Rahim |
| Mosaddek Hossain | 71.0 | Nasir Hossain |

| | | | |
|---|---|---|---|
| Sabbir Rahman | 82.88 | Sabbir Rahman | |
| Shuvagata Home | 90.83 | Mustafizur Rahman | |
| Mahmudullah | 82.8 | Mahmudullah | |
| Shakib Al- Hasan | 88.0 | Shakib Al- Hasan | |
| Alok Kapali | 79.76 | Arafat Sunny | |
| Mehedi Hasan | 75.41 | Jubair Hossain | |
| Mashrafee Murtaza | 81.92 | Mashrafe Murtaza | |
| Taskin Ahmed | 79.99 | Taskin Ahmed | |
| Al-Amin Hossain | 78.77 | Al-Amin Hossain | |

and finally our ideal squad of Bangladesh National Cricket team using statistical analysis is shown in Table 10.

Table 10: An Ideal Squad for Bangladesh National Cricket team

| Player Name |
|---|
| Soumya Sarkar |
| Tamim Iqbal |
| Imrul Kayes |
| Salman Hossain |
| Mushfiqur Rahim |
| Mosaddek Hossain |
| Sabbir Rahman |
| Shuvagata Home |
| Mahmudullah |
| Shakib Al- Hasan |
| Alok Kapali |
| Mehdi Hasan |
| Mashrafee Murtaza |
| Taskin Ahmed |
| Al-Amin Hossain |